\documentstyle[preprint,revtex,epsf]{aps}

\tightenlines

\begin{document}
\draft
\centerline{Phys. Rev. E {\bf 65}, 1 February (2002)}
\vskip 20 mm
\preprint{}

\begin{title}
Exact amplitude ratio and finite-size corrections\\
for the $M \times  N$ square lattice Ising model
\end{title}

\author{N.Sh. Izmailian$^{1,2}$\quad  and   Chin-Kun Hu$^{1,3,*}$}

\begin{instit}
$^1$ \normalsize Institute of Physics, Academia Sinica,
Nankang, Taipei 11529, Taiwan \\
\end{instit}
\begin{instit}
$^2$ \normalsize Yerevan Physics Institute, Alikhanian Br. 2, 375036
Yerevan, Armenia
\end{instit}

\begin{instit}
$^3$ \normalsize Department of Physics, National Dong Hwa University, Hualien
 97401, Taiwan
\end{instit}

\begin{abstract}
Let $f$, $U$ and $C$ represent, respectively, the free energy,
the internal energy and the specific heat of the
critical Ising model on the $M \times N$ square lattice with periodic
boundary conditions,
and  $f_{\infty}$ represents $f$ for fixed $M/N$ and $N \to \infty$.
We find that $f$, $U$ and $C$ can be written as:
 $N(f - f_{\infty})=\sum_{i=1}^{\infty}{f_{2i-1}}/{N^{2 i-1}}$,
$U=-\sqrt{2}+\sum_{i=1}^{\infty}u_{2i-1}/N^{2 i-1}$  and
$C=8 \ln{N}/\pi+\sum_{i=0}^{\infty}c_i/N^{i}$, i.e.
$N f$ and $U$ are odd functions of $N^{-1}$.
We also find that $u_{2i-1}/c_{2i-1} = 1/\sqrt{2}$ and
$u_{2 i}/c_{2 i} = 0$ for $1 \le i < \infty$
and obtain closed form expressions for $f$, $U$, and $C$ up to
orders $1/N^5$, $1/N^5$, and $1/N^3$, respectively,
which implies an analytic equation for $c_5$.
\end{abstract}

\vskip 5 mm

\vskip 10 cm

\newpage

\vskip 8 mm

\section{Introduction}
The Ising model has been used to represent critical phenomena in
ferromagnets, binary alloys, binary fluids, gas-liquid mixture, etc
and is perhaps the most widely studied model of critical phenomena
\cite{stanley71}. For analyzing the simulation
or experimental data of finite critical systems \cite{binder92},
it is useful to appeal to theories of finite-size corrections \cite{ff69}
and finite-size scaling \cite{fss}. Such theories
have attracted much attention in recent years
\cite{huetal,hcik99,queiroz,sokal,ih00} because of the fast advance in
computers' computing power and algorithms for simulating or analyzing data.
Theories of finite-size effects
and of finite-size scaling in general
have been most successful in deriving critical and noncritical
properties of infinite systems from those of their finite or partially
finite counterparts.
Finite-size corrections and finite-size scaling for the $M \times N$ square
lattice Ising model are of particular interest because the Ising model
is very popular and such system is
usually used to test the efficiency of algorithms for studying critical
systems \cite{test}. In the present paper, we
present new analytic results for finite-size effects in the Ising model
on a large $M \times N$ square lattice at the critical point.

Finite-size scaling is the basis of the powerful phenomenological
renormalization group method \cite{wegner,night}. In the two-dimensional
Ising model the finite-size effect on the renormalization transformation
has been demonstrated to be rather benign \cite{dudek}, and the effects due
to convergence to the fixed point and finite-size are clearly distinguished
\cite{blote}. The finite-size scaling theory predicts that near the
critical point the singular part of the thermodynamic quantity
of a finite system, say $\it Q_{s}$, has the scaling form
\begin{equation}
\it Q_{s} = L^{y_{\it Q}}Y_{\it Q}(L/\xi_{\infty}),
\label{anzats}
\end{equation}
where $L$ is system linear size, $\xi_{\infty}$ is the correlation length of
the bulk system, $y_{\it Q}$ is a critical exponent and $Y_{\it Q}$ is the
scaling function. The scaling ansatz mentioned above ignores the possible
logarithmic corrections. In the case of planar Ising model, which displays
a logarithmic singularities in the specific heat behavior due to a relation
between scaling exponents in the renormalization group theory
\cite{wegner}, the
scaling form (\ref{anzats}) must then be replaced by a more general form
\cite{fss,privmanrud,helen,guo}
\begin{equation}
\it Q_{s} = L^{y_{\it Q}}Y_{\it Q}(L/\xi_{\infty})+
L^{y_{\it Q}}\ln{L}\; X_{\it Q}(L/\xi_{\infty}),
\label{anzats1}
\end{equation}
which in the case of the specific heat ($C$) becomes
\begin{equation}
\it C_{s} = Y_{C}(L/\xi_{\infty})+
\ln{L}\; X_{C}(L/\xi_{\infty}).
\label{anzats2}
\end{equation}
The results of this paper to be presented below show that the leading term
of $C_s$ is $8\ln{L}/\pi$, all other
finite-size corrections to the specific
heat are always integer powers of $L^{-1}$, which also imply that the
scaling function $X_{C}$ in Eq. (\ref{anzats2}) is constant and equal to
$8/\pi$. Very recently, Caselle {\it et al.} \cite{caselle} have shown that
this result can be predicted by conformal field theory under a number of
general conjectures.

The relevance of the finite-size properties to the conformal field theory is
another source of interest. Discussion of general
properties of non-universal corrections to finite size scaling and their
relation to irrelevant operators in conformal field theory can be found
in \cite{Henkel}. On the basis of conformal invariance,
the asymptotic finite-size
scaling behavior of the critical free energy $f_N$ per site and the inverse
correlation length $\xi_N^{-1}$ of a $N \times \infty$ system is found to
be \cite{affleck}
\begin{equation}
\lim_{N \to \infty} {N^2(f_N-f_{\infty})}=\frac{c \pi}{6},
\label{I2}
\end{equation}
\begin{equation}
\lim_{N \to \infty} {N \xi_N^{-1}} =
2 \pi x,
\label{I1}
\end{equation}
where $f_{\infty}$ is the free energy of
the bulk system, $c$ is the conformal anomaly number and $x$ is the
scaling dimension.
The corrections to Eqs. (\ref{I2}) and (\ref{I1})  can be calculated by
the means of a perturbated conformal field theory \cite{cardy86,zamol87} and
can be expressed in terms of the universal structure
constants $(C_{nln})$ of the operator product expansion \cite{cardy86}.
Quite recently, Izmailian and Hu \cite{ih00} studied the finite-size
correction terms for the free energy and the inverse correlation length of
critical Ising model on $N \times \infty$ lattices and obtained a new set of
the universal amplitude ratios for the coefficients in
the free energy and the inverse correlation length expansions. It was shown
that such results could be understood from a perturbated conformal field
theory.

Based on Onsager's solution, explicit calculations of the specific heat
finite-size scaling behavior have been reported by Ferdinand and
Fisher \cite{ff69} and by Kleban and Akinci \cite{kleban83}.
In 1969,  Ferdinand and Fisher \cite{ff69} first studied finite-size
corrections for a critical
Ising model on $M \times N$ square lattices with periodic
boundary conditions.  They gave explicit expressions for the critical
 free energy $f$, internal energy $U$, and specific
heat $C$ per lattice site for a fixed $\xi=M/N$ and large $N$
up to orders $1/N^2$, $1/N$, and $1/N$, respectively:

\begin{eqnarray}
f &=&  f_{\infty} + \frac{1}{\xi N^2}\left[
\ln{(\theta_2+\theta_3+\theta_4)}-\frac{1}{3}\ln{(4\theta_2 \theta_3
\theta_4)}
\right] +O\left(\frac{1}{N^3}\right)
\label{freeen}\\
U &=& -\sqrt{2} - \frac{1}{N}
\frac{2{\theta}_2{\theta}_3{\theta}_4}{{\theta}_2+{\theta}_3+{\theta}_4}
+O\left(\frac{1}{N^2}\right)
\label{internal}\\
C&=&\frac{8}{\pi}\ln{N}+\frac{8}{\pi}\left(\ln
{\frac{2^{5/2}}{\pi}}+C_E- \frac{\pi}{4}\right)
\nonumber\\
&-&\frac{16}{\pi}\frac{\sum_{i=2}^4{\theta}_i \ln{{\theta}_i}}
{{\theta}_2+{\theta}_3+{\theta}_4}-4 \xi \left(\frac{
{\theta}_2{\theta}_3{\theta}_4}{{\theta}_2+{\theta}_3+{\theta}_4}\right)^2
\label{spheat}\\
&-&2\sqrt{2}\frac{
{\theta}_2{\theta}_3{\theta}_4}{{\theta}_2+{\theta}_3+{\theta}_4}\frac{1}{N}
+O((\ln{N})^3/N^2)
\nonumber
\end{eqnarray}
where $\theta_i=\theta_i(0,q)$ $(i = 2, 3, 4)$ is elliptic theta
functions of modulus
$q=e^{-\pi\xi}$, $C_E$ is the Euler constant and $f_{\infty}$ is the free
energy in the thermodynamic limit $M,N \to \infty$.

In 1983, Kleban and Akinci \cite{kleban83} gave a very accurate and relatively
simple approximate closed form expression for leading specific-heat
correction term which results
from retaining  only the two largest eigenvalues of
the transfer matrix. This approximation is already good at $\xi = 1$ and
becomes exponentially better with increasing $\xi$, and they interpreted
their results in terms of domain-wall energies.

In this paper we study the same system as \cite{ff69} and find that
$f$, $U$ and $C$ can be written as:
 $N(f - f_{\infty})=\sum_{i=1}^{\infty}{f_{2i-1}}/{N^{2 i-1}}$,
$U=-\sqrt{2}+\sum_{i=1}^{\infty}u_{2i-1}/N^{2 i-1}$  and
$C=8 \ln{N}/\pi+\sum_{i=0}^{\infty}c_i/N^{i}$, i.e.
$N f$ and $U$ are odd functions of $N^{-1}$.
We also find that $u_{2i-1}/c_{2i-1} = 1/\sqrt{2}$ and
$u_{2 i}/c_{2 i} = 0$ for $1 \le i < \infty$
and obtain analytic equations for $f$, $U$, and $C$ up to
orders $1/N^5$, $1/N^5$, and $1/N^3$, respectively,
which implies an analytic equation for $c_5$.

We have also shown that
Kleban and Akinci approximation is in excellent agreement with our exact
results for the leading correction terms of the free
energy ($f_1$), the internal energy ($u_1$) and the specific heat
($c_0, c_1$) at $\xi \ge 1$.  For the next correction terms the error
introduced by the
two-eigenvalues approximation is maximum at
$\xi = 1 \; (M=N)$. With increasing $\xi$ the exact and approximate values
approach exponentially and approximation becomes already good
at $\xi = 1.65$ for the correction terms $f_3, u_3, c_2, c_3$ and at
$\xi = 1.85$ for the correction terms $f_5, u_5$.

This paper is organized as follows. In Sec. II, we write the free
energy $f$, the internal energy $U$, and the specific heat $C$
of the Ising model
in terms of $P_1$, $P_2$, $P_3$, $P_4$, $Q_1$, $Q_2$, and $Q_3$
defined in this section. In Sec. III, we present asymptotic expansions
for $f$, $U$, and $C$. In Sec. IV, we discuss some problems
for further studies. Some mathematical details used in the derivation
of equations in Sec. III are given in Appendix A.

\vskip 8 mm

\section{Ising model}

Consider an Ising ferromagnet on an $M \times N$ lattice
with periodic boundary conditions ({\it i.e.} a torus).
The Hamiltonian of the system is
\begin{equation}
\beta H=-J\sum_{<ij>} s_i s_j,
\label{hamilt}
\end{equation}
where $\beta = (k_BT)^{-1}$, the Ising spins $s_i=\pm 1$ are located at
the sites of the lattice and the summation goes over all
nearest-neighbor pairs of the lattice.
The partition function $Z_{MN}(T)$ of a finite  $M \times N$ square Ising
lattice wrapped on a torus can be written as
\begin{equation}
Z_{MN}(T)=\frac{1}{2}\left(2\sinh{2 J}\right)^{\frac{1}{2}MN}\sum_{i=1}^4Z_i,
\label{partition}
\end{equation}
where the partial partition functions $Z_i$ are defined by
\begin{eqnarray}
Z_1&=&\prod_{r=0}^{N-1}2\cosh{\frac{M}{2}\gamma_{2 r+1}}=
P_1\exp{\left[\frac{M}{2}\sum_{r=0}^{N-1}\gamma_{2 r+1}\right]},
\label{Z1}\\
Z_2&=&\prod_{r=0}^{N-1}2\sinh{\frac{M}{2}\gamma_{2 r+1}}=
P_2\exp{\left[\frac{M}{2}\sum_{r=0}^{N-1}\gamma_{2 r+1}\right]},
\label{Z2}\\
Z_3&=&\prod_{r=0}^{N-1}2\cosh{\frac{M}{2}\gamma_{2 r}}=
P_3\exp{\left[\frac{M}{2}\sum_{r=0}^{N-1}\gamma_{2 r}\right]}(1+
e^{-M \gamma_0}),
\label{Z3}\\
Z_4&=&\prod_{r=0}^{N-1}2\sinh{\frac{M}{2}\gamma_{2 r}}
=P_4\exp{\left[\frac{M}{2}\sum_{r=0}^{N-1}\gamma_{2 r}\right]}(1-
e^{-M \gamma_0}),
\label{Z4}
\end{eqnarray}
with
\begin{eqnarray}
P_1&=&\prod_{r=0}^{N-1}(1+e^{- M \gamma_{2r+1}}),
\qquad  \qquad
P_2=\prod_{r=0}^{N-1}(1-e^{- M \gamma_{2r+1}}),
\nonumber \\
P_3&=&\prod_{r=1}^{N-1}(1+e^{- M \gamma_{2r}}),
\qquad \qquad \quad
P_4=\prod_{r=1}^{N-1}(1-e^{- M \gamma_{2r}}),
\label{P}
\end{eqnarray}
and $\gamma_r$  is implicitly given by

\begin{eqnarray}
\cosh{\gamma_r}=\frac{\cosh^2{2 J}}{\sinh{2 J}}-\cos{\frac{r \pi}{N}}.
\label{gammasq}
\end{eqnarray}
At the critical point $J_c$ of the sq lattice Ising model,
where $J_c=\frac{1}{2}\ln{(1+\sqrt{2})}$,
one then obtains
\begin{equation}
\gamma^{(cr)}_r= 2 \psi_{sq} \left(\frac{r \pi}{2 N}\right), \qquad
\mbox{with}
\qquad \psi_{sq}(x)=\ln{\left(\sin{x}+\sqrt{1+\sin^2{x}}\right)}.
\label{gammacrit}
\end{equation}
The free energy, the internal energy per spin and the specific heat per spin
 can be obtained from the partition function $Z_{MN}$
\begin{eqnarray}
f&=&\frac{1}{MN} \ln{Z_{MN}}=
\frac{1}{2}\ln{2\sinh{2 J}}+\frac{1}{M N}\ln{\frac{1}{2}}
\left(\sum_{i=1}^4Z_i \right),
\label{freeenergy1}\\
U&=&-\frac{1}{MN} \frac{d}{d J}\ln{Z_{MN}}=
-\coth{2 J}-\frac{1}{MN}\left(\sum_{i=1}^4Z_i'\right)/\left(\sum_{i=1}^4Z_i\right),
\label{internalenergy1}\\
C &=&\frac{1}{MN} \frac{d^2}{d J^2}\ln{Z_{MN}}
\label{spheat1}\\
&=&
-\frac{2}{\sinh^2 {2 J}}+\frac{1}{MN}\left\{\left(
\sum_{i=1}^4Z_i''\right)/\left(\sum_{i=1}^4Z_i\right)-
\left[\left(\sum_{i=1}^4Z_i'\right)/\left(\sum_{i=1}^4Z_i\right)\right]^2\right\},
\nonumber
\end{eqnarray}
where the primes denote differentiation with respect to $J$.
At the critical point ($T=T_c$) the partial partition functions $Z_i$ and
their first and second derivatives are given by
\begin{eqnarray}
Z_1 &=& P_1 e^A, \qquad Z_2 = P_2 e^A, \qquad Z_3 = 2 P_3 e^B, \qquad Z_4 = 0;
\label{Zic}\\
Z_1' &=& 0, \qquad \quad \: Z_2' = 0, \quad \qquad \, Z_3' = 0, \quad \quad
\qquad \! Z_4' = 4 M P_4 e^B;
\label{Zi'c}\\
\frac{Z_1''}{M N} &=& Q_1 Z_1, \quad \frac{Z_2''}{M N} = Q_2 Z_2, \quad
\frac{Z_3''}{M N} = Q_3 Z_3, \quad Z_4'' = - \sqrt{2} Z_4',
\label{Zi''c}
\end{eqnarray}
where
\begin{equation}
A = \frac{M}{2}\sum_{r=0}^{N-1}\gamma^{(cr)}_{2 r+1},
\qquad  B = \frac{M}{2} \sum_{r=0}^{N-1}
\gamma^{(cr)}_{2 r},
\label{AB}
\end{equation}
\begin{eqnarray}
Q_1&=& \frac{1}{2 N}\sum_{r=0}^{N-1}{}\gamma_{2 r+1}''^{(cr)}
\tanh{\frac{M \gamma_{2 r+1}^{(cr)}}{2}},
\label{Q1}\\
Q_2&=& \frac{1}{2 N}\sum_{r=0}^{N-1}{}\gamma_{2 r+1}''^{(cr)}
\coth{\frac{M \gamma_{2 r+1}^{(cr)}}{2}},
\label{Q2}\\
Q_3&=& 4 \xi +\frac{1}{2 N}\sum_{r=1}^{N-1}{}\gamma_{2 r}''^{(cr)}
\tanh{\frac{M \gamma_{2 r}^{(cr)}}{2}},
\label{Q3}
\end{eqnarray}
and $P_i$ are given by Eq. (\ref{P}) with $\gamma_r=\gamma_r^{(cr)}$ and
$\gamma_r''^{(cr)}$ denote the second derivative of $\gamma_r$ with respect
to $J$ at the critical point $J = J_c$.
Then the exact expression for the free energy,
the internal energy and the specific heat of a
finite Ising model  at critical point ($T = T_c$) can be written as
\begin{eqnarray}
f &=& \frac{1}{2} \ln{2} + \frac{1}{M N} A + \frac{1}{M N}
\ln{\frac{P_1+P_2+2 P_3 \exp{(B - A)}}{2}},
\label{freeen2}\\
U &=& - \sqrt{2} - \frac{4}{N} \frac{P_4}{2 P_3 +(P_1+P_2)\exp{(A - B)}},
\label{intern2} \\
C &=& \sqrt{2}  U -
\xi \left(\frac{4 P_4}{2 P_3 +(P_1+P_2)\exp{(A - B)}}\right)^2
\nonumber\\
&+&\frac{2 Q_3 P_3+(Q_1 P_1+Q_2 P_2)\exp{(A - B)}}
{2 P_3 +(P_1+P_2)\exp{(A - B)}}.
\label{spheat2}
\end{eqnarray}

\vskip 8 mm

\section{Asymptotic expansions}

We consider only  sequences of lattices in which $\xi=M/N$
remains positive and finite as the thermodynamic limit $M, N \to \infty$
is approached. Using Taylor's theorem we find that $M \gamma_r$ is
even function of $1/N$ at the critical point
\begin{equation}
M \gamma_r^{(cr)} =\sum_{i=0}^{\infty}\frac{a_i}{N^{2 i}}=
 \pi \xi r - \frac{\pi^3 \xi}{12} \frac{r^3}{N^2}+
\frac{\pi^5 \xi}{96} \frac{r^5}{N^4}+....
\label{gammar}
\end{equation}
Using Euler-Maclaurin summation formula \cite{Hardy} we can expand $A$ and
$B$ up to arbitrary order
\begin{eqnarray}
A &=& \frac{M N}{\pi}\int_0^{\pi} \psi_{sq}(x) d x
 + M \sum_{k=1}^{\infty}
\frac{2 B_{2 k}}{(2 k)!}(2^{2 k -1}-1)\psi_{sq}^{(2 k-1)}(0)
\left(\frac{\pi}{2 N}\right)^{2 k-1}
\nonumber \\
&=& \frac{2 G}{\pi} M N +
\frac{\pi \xi}{12}+\frac{7 \pi^3 \xi}{1440} \frac{1}{N^2}+
\frac{31 \pi^5 \xi}{24192} \frac{1}{N^4}
+ \frac{10033 \pi^7 \xi}{9676800}\frac{1}{N^6}+...,
\label{2f}\\
B &=&\frac{M N}{\pi}\int_0^{\pi} \psi_{sq}(x) d x
 -M \sum_{k=1}^{\infty}
\frac{2 B_{2 k}}{(2 k)!}\left(\frac{\pi}{N}\right)^{2 k-1}
\psi_{sq}^{(2 k-1)}(0)
\nonumber\\
&=& \frac{2 G}{\pi} M N -
\frac{\pi \xi}{6}-\frac{\pi^3 \xi}{180} \frac{1}{N^2}-
\frac{\pi^5 \xi}{756} \frac{1}{N^4}
- \frac{79 \pi^7 \xi}{75600}\frac{1}{N^6}-\dots,
\label{sum1f}
\end{eqnarray}
where $B_{2 i}$ are the Bernoulli
numbers and $G=0.915965...$ is Catalan's constant.

Let us now evaluate the products $P_i$ for $i = 1, 2, 3, 4$. It is easy to
see from Eqs. (\ref{P}) and (\ref{gammar}) that $P_i$ contains only even
power
of $1/N$
\begin{equation}
P_i=\pi_i(0,\xi)\left(1+\sum_{j=1}^{\infty}\frac{p_{ij}}{N^{2 j}}\right)
\qquad (i = 1, 2, 3, 4),
\label{Pi}
\end{equation}

\begin{eqnarray}
P_1&=&\pi_1(0,\xi)\left(1+\frac{p_{11}}{N^2}+\frac{p_{12}}{N^4}+...
\right),
\quad
P_2=\pi_2(0,\xi)\left(1+\frac{p_{21}}{N^2}+\frac{p_{22}}{N^4}+...
\right),
\nonumber \\
P_3&=&\pi_3(0,\xi)\left(1+\frac{p_{31}}{N^2}+\frac{p_{32}}{N^4}+...
\right),
\quad
P_4=\pi_4(0,\xi)\left(1+\frac{p_{41}}{N^2}+\frac{p_{42}}{N^4}+...
\right),
\nonumber
\end{eqnarray}
with
$$
\pi_1(0,\xi)=\frac{\theta_3}{\theta_0}, \quad
\pi_2(0,\xi)=\frac{\theta_4}{\theta_0}, \quad
\pi_3(0,\xi)=\frac{\theta_0^2}{\theta_3 \theta_4}, \quad
\pi_4(0,\xi)=\theta_0^2 , \quad
e^{\frac{\pi \xi}{4}}=\frac{2 \theta_0^3}{\theta_2 \theta_3 \theta_4},
$$
where $\theta_i = \theta_i(0,q)$ is elliptic theta functions of modulus
$q = e^{-\pi\xi}$. The explicit expressions for the coefficients $p_{i1}$ and
$p_{i2}$ for $i = 1, 2, 3, 4$ are given in Appendix {\bf A}.

One readily sees from Eqs. (\ref{freeen2}), (\ref{intern2}) and
(\ref{2f}) - (\ref{Pi}) that
the finite-size estimates of the free energy ($N f$) and the internal
energy ($U$) must be odd functions of $N^{-1}$.

\begin{eqnarray}
N(f - f_{\infty}) &=& \sum_{i=1}^{\infty}{\frac{f_{2 i-1}}{N^{2 i-1}}},
\label{free} \\
U &=& - \sqrt{2} + \sum_{i=1}^{\infty}{\frac{u_{2 i-1}}{N^{2 i-1}}}.
\label{energy}
\end{eqnarray}

Substituting Eqs. (\ref{2f}) - (\ref{Pi}), (\ref{pi1fin}) and (\ref{pi2fin})
 in Eqs. (\ref{freeen2}) and (\ref{intern2}) we can write the
expansions of the free energy ($N f$) and the internal energy ($U$) at
the critical point $(T=T_c)$ up to $1/N^5$ order. The final result is
\begin{eqnarray}
N(f-f_{\infty})&=&\frac{\xi^{-1}}{N}\left[
\ln{(\theta_2+\theta_3+\theta_4)}-\frac{1}{3}\ln{(4\theta_2 \theta_3
\theta_4)}
\right] \nonumber\\
&-&\frac{\pi^3}{N^3}\frac{
8{\theta}_2 {\theta}_3 {\theta}_4
\left[
{\theta}_3^3({\theta}_2^3+{\theta}_4^3)-{\theta}_2^3 {\theta}_4^3
\right]-7(
{\theta}_2^9+{\theta}_3^9+{\theta}_4^9)
}
{
1440({\theta}_2+{\theta}_3+{\theta}_4)
} \label{freeenfin}\\
&+& \frac{1}{N^5}\frac{\xi K^8}{189 \pi^2}\frac{f_{51}+
\left(\frac{E'}{K'} - \frac{E}{K}\right)f_{52}}
{{\theta}_2+{\theta}_3+{\theta}_4}+O\left(\frac{1}{N^7}\right),
\nonumber
\end{eqnarray}
with
\begin{eqnarray}
f_{52}&=&\theta_2 (-32+48 k^2-78 k^4+31 k^6)
+\theta_3 (31-78 k^2+48 k^4-32 k^6)
\nonumber\\
&+&\theta_4 (31-15 k^2-15 k^4+31 k^6)
\label{f52}\\
f_{51} &=& \theta_2 (-32+80 k^2-38 k^4+21 k^6)
+\theta_3 (31-88 k^2+88 k^4)
\nonumber\\
&+&\theta_4 (31-67 k^2+25 k^4-21 k^6)
\nonumber\\
&+&\frac{21}{8}\frac{\theta_3\theta_4 k^4 (1+{k'}^2)^2
+\theta_2\theta_3 {k'}^4(1+k^2)^2
+\theta_2\theta_4({k'}^2-k^2)^2}{{\theta}_2+{\theta}_3+{\theta}_4},
\label{f51}
\end{eqnarray}
and
\begin{eqnarray}
U&=&-\sqrt{2}-\frac{2}{N}
\frac{{\theta}_2{\theta}_3{\theta}_4}{{\theta}_2+{\theta}_3+{\theta}_4}
+\frac{2}{N^3}\frac{
{\theta}_2{\theta}_3{\theta}_4({\theta}_2^9+{\theta}_3^9+{\theta}_4^9)}{
({\theta}_2+{\theta}_3+{\theta}_4)^2}\frac{\pi^3 \xi}{96}
\label{internfin}\\
&+&\frac{2}{N^5}\frac{
{\theta}_2{\theta}_3{\theta}_4}{
({\theta}_2+{\theta}_3+{\theta}_4)^2}\frac{\xi^2 K^8}{3 \pi^2}
\left[U_{51}+\left(\frac{E'}{K'} - \frac{E}{K}\right)U_{52}\right]
+O\left(\frac{1}{N^7}\right)
\nonumber
\end{eqnarray}
with
\begin{eqnarray}
U_{52}&=&\theta_3 ({k'}^2-k^2)-\theta_2  k^4(1+{k'}^2)+\theta_4 {k'}^4
(1+k^2),
\label{U52}\\
U_{51}&=&\theta_3 \frac{23-64 k^2+64 k^4}{24}
+\theta_2 k^4\frac{16+8 k^2-k^4}{24}
+\theta_4 (1-k^2)^2\frac{23-6 k^2-k^4}{24}
\nonumber\\
&+&\frac{\theta_3\theta_4 k^4(1+{k'}^2)^2
+\theta_2\theta_3 {k'}^4(1+k^2)^2
+\theta_2\theta_4({k'}^2-k^2)^2}{12 ({\theta}_2+{\theta}_3+{\theta}_4)},
\label{U51}
\end{eqnarray}
where $f_{\infty} = - 0.5 \ln{2} - 2 G/{\pi}$ and ${\theta}_2, {\theta}_3,
{\theta}_4$ are elliptic function
\begin{equation}
{\theta}_2=\sqrt{\frac{2 k K(k)}{\pi}}, \qquad
{\theta}_3=\sqrt{\frac{2 K(k)}{\pi}}, \qquad
{\theta}_4=\sqrt{\frac{2 k' K(k)}{\pi}},
\label{thetai}
\end{equation}
with $K(k)$ and $E(k)$ the elliptic integrals of the
first and second kind, respectively. For simplicity we denote $K \equiv K(k)$,
$K' \equiv K'(k)$,
$E \equiv E(k)$ and $E' \equiv E'(k)$.

It is interesting to compare our results for the free energy and the
internal energy with Kleban and
Akinci two eigenvalues approximation. Keeping only two largest eigenvalues
$\lambda_0$ and $\lambda_1$ of the transfer matrix, the partition function of
the Ising model can be written as
\begin{equation}
Z_{MN} = \lambda_0^M+\lambda_1^M
\label{kleban}
\end{equation}
with
\begin{eqnarray}
\lambda_0 &=&  (2 \sinh 2 J)^{N/2}
\exp{\left(\frac{1}{2}\sum_{r=0}^{N-1} \gamma_{2 r +1}\right)},
\label{lambda0}\\
\lambda_1 &=& (2 \sinh 2 J)^{N/2}
\exp{\left(\frac{1}{2}\sum_{r=1}^N \gamma_{2 r}\right)},
\label{lambda1}
\end{eqnarray}
where $\gamma_{k}$ is implicitly given by Eq. (\ref{gammasq}).

To write the critical free energy $f$ and critical internal energy
$U$ in the form of Eqs. (\ref{free}) and
(\ref{energy}),
we must evaluate Eqs. (\ref{lambda0}) and (\ref{lambda1}) asymptotically.
These sums can be handled by using the Euler-Maclaurin summation formula
\cite{Hardy}. After a straightforward calculation, we have obtained
\begin{eqnarray}
N(f - f_{\infty}) &=&\frac{f_1^{(app)}}{N}+\frac{f_3^{(app)}}{N^{3}}+
\frac{f_5^{(app)}}{N^{5}}+... ,
\label{freekleban} \\
U &=& - \sqrt{2} +\frac{u_1^{(app)}}{N}+\frac{u_3^{(app)}}{N^{3}}+
\frac{u_5^{(app)}}{N^{5}}+...
\label{energykleban}
\end{eqnarray}
with
\begin{eqnarray}
f_1^{(app)}&=&-\frac{\pi}{12}-\frac{1}{\xi}\ln{\left(1+e^{-\pi \xi/4}\right)}
\label{f1app}\\
f_3^{(app)}&=&\frac{\pi^3}{2880}\left[1-15 \tanh{\left(\pi \xi/8\right)}
\right]
\label{f2app}\\
f_5^{(app)}&=& \frac{\pi^5}{48384}\left[1-63 \tanh{\left(\pi \xi/8\right)}
\right] - \frac{\pi^6 \xi}{73728}{\rm sech}^2{\left(\pi \xi/8\right)}
\label{f3app}\\
u_1^{(app)}&=&-1+\tanh{\left(\pi \xi/8\right)}
\label{u1app}\\
u_3^{(app)}&=& \frac{\pi^3 \xi}{192}{\rm sech}^2{\left(\pi \xi/8\right)}
\label{u2app}\\
u_5^{(app)}&=& \frac{\pi^5 \xi}{768} {\rm sech}^2{\left(\pi \xi/8\right)}
-\frac{\pi^6 \xi^2}{36864}{\rm sech}^2{\left(\pi \xi/8\right)}
\tanh{\left(\pi \xi/8\right)}
\label{u3app}
\end{eqnarray}
The expressions of the coefficients given by Eqs. (\ref{f1app}) -
(\ref{u3app}) are much simpler than their exact counterparts given by
Eqs. (\ref{freeenfin}) - (\ref{U51}). Nevertheless, one can see from
Figs. 1 and 2 that two-eigenvalues approximation proposed by Kleban and
Akinci  is already good at $\xi = 1$ for the leading
corrections terms in the free energy ($f_1$) and the internal
energy ($u_1$) and becomes exponentially better with increasing
$\xi$. The error introduces by the
two-eigenvalues approximation is maximum at
$\xi = 1 \; (M=N)$. With increasing $\xi$ the exact and approximate values
approach exponentially and approximation becomes already good
at $\xi = 1.65$ for the correction terms $f_3, u_3$ and at
$\xi = 1.85$ for the correction terms $f_5, u_5$. We consider the case
$\xi \ge 1$ only. By symmetry, the same results hold for $\xi'=1/\xi \le 1$.

To calculate the specific heat we must also evaluate asymptotically
the sums appearing in the expression (\ref{spheat2}) for $C$, namely
$Q_1, Q_2$ and $Q_3$. Since
the
analysis follows the same general lines as in the cases of the free energy and
the internal energy,
we will not present the details of calculations and we quote here
only results, namely, at the critical point $T=T_c$ the asymptotic expansion
of the sums $Q_1, Q_2$ and $Q_3$ can be written as
\begin{eqnarray}
Q_i &=& \frac{8}{\pi}\ln{N} + \sum_{j=0}^{\infty}\frac{q_{ij}}{N^{2 j}}
\qquad \mbox{for} \qquad (i = 1, 2, 3)
\label{Qi}
\end{eqnarray}
where $q_{i0}$ and $q_{i1}$ (for $i = 1, 2, 3$) are given by
\begin{eqnarray}
q_{10} &=& \frac{8}{\pi}(C_E+\ln\frac{2^{5/2}}{\pi}-2 \ln{\theta_3})
\label{q10}\\
q_{20} &=& \frac{8}{\pi}(C_E+\ln\frac{2^{5/2}}{\pi}-2 \ln{\theta_4})
\label{q20}\\
q_{30} &=& \frac{8}{\pi}(C_E+\ln\frac{2^{5/2}}{\pi}-2 \ln{\theta_2})
\label{q30}\\
q_{11} &=& -\frac{8 K^4 \xi}{9 \pi^2}\left[1+(1-2 k^2)
\left(\frac{E'}{K'} - \frac{E}{K}\right)\right],
\label{q11}\\
q_{21} &=& -\frac{8 K^4 \xi}{9 \pi^2}\left[1-3 k^2+(1+ k^2)
\left(\frac{E'}{K'} - \frac{E}{K}\right)\right],
\label{q21}\\
q_{31} &=& \frac{8 K^4 \xi}{9 \pi^2}\left[2-3 k^2+(2- k^2)
\left(\frac{E'}{K'} - \frac{E}{K}\right)\right].
\label{q31}
\end{eqnarray}
It is easy to see from Eqs. (\ref{intern2}), (\ref{spheat2}), (\ref{Pi}) and
(\ref{Qi}) that the asymptotic expansion of the specific heat,
can be written as
\begin{equation}
C=\frac{8}{\pi} \ln{N}+\sum_{i=0}^{\infty}c_i/N^{i}.
\label{C}
\end{equation}
Except for the leading term, all other corrections in the asymptotic expansion
of the specific heat are proportional to $1/N^{i}$,
without multiplicative logarithms. This
result imply immediately that scaling function $X_C$ in Eq. (\ref{anzats2})
is constant and equal to $8/\pi$.

It is also clear that the contribution to odd $(N^{-2 i -1})$ order in
the specific heat expansion give only first term in right-hand side of the
Eq. (\ref{spheat2}). Thus, we can obtained immediately that the
ratio $u_{2 i+1}/c_{2 i+1}$ of subdominant $(N^{-2 i -1})$ finite-size
corrections
term in the internal energy and the specific heat expansions are constant,
namely,
\begin{equation}
u_{2 i+1}/c_{2 i+1} = 1/\sqrt{2}
\label{ratio}
\end{equation}
as well that
$u_{2 i}/c_{2 i} = 0$ for $1 \le i < \infty$.

Let us now evaluate the first few terms in the specific heat expansion.
Substituting Eqs. (\ref{2f}) - (\ref{Pi}),
(\ref{Qi}), (\ref{q10}) - (\ref{q31}), (\ref{pi1fin}) and (\ref{pi2fin})
in Eq. (\ref{spheat2}) we have finally
obtained the expansion of the specific heat ($C$) at the critical point
$(T=T_c)$
\begin{eqnarray}
C&=&\frac{8}{\pi}\ln{N}+\frac{8}{\pi}
\left(\ln
{\frac{2^{5/2}}{\pi}}+C_E- \frac{\pi}{4}\right)-4 \xi \left(\frac{
{\theta}_2{\theta}_3{\theta}_4}{{\theta}_2+{\theta}_3+{\theta}_4}\right)^2
\nonumber\\
&-&\frac{16}{\pi}\frac{\sum_{i=2}^4{\theta}_i \ln{{\theta}_i}}
{{\theta}_2+{\theta}_3+{\theta}_4}-2\sqrt{2}\frac{
{\theta}_2{\theta}_3{\theta}_4}{{\theta}_2+{\theta}_3+{\theta}_4}
\frac{1}{N}+\frac{c_2}{N^2}
\label{spheatfin}\\
&+& \frac{1}{N^3}\frac{\pi^3 \xi}{24\sqrt{2}}
\frac{{\theta}_2 {\theta}_3 {\theta}_4
({\theta}_2^9+{\theta}_3^9+{\theta}_4^9)}
{({\theta}_2+{\theta}_3+{\theta}_4)^2}
+O\left(\frac{1}{N^4}\right)+O\left(\frac{1}{N^5}\right),
\nonumber
\end{eqnarray}
with
\begin{eqnarray}
c_2 &=& \frac{\pi^3 \xi^2}{12}\;
\frac{{\theta}_2^2 {\theta}_3^2 {\theta}_4^2
({\theta}_2^9+{\theta}_3^9+{\theta}_4^9)}
{({\theta}_2+{\theta}_3+{\theta}_4)^3}
+\frac{\pi^2 \xi}{9}\;
\frac{{\theta}_3^4 {\theta}_4^4(2{\theta}_2-{\theta}_3-{\theta}_4)}
{{\theta}_2+{\theta}_3+{\theta}_4}
\nonumber\\
&-&\frac{\pi^2 \xi}{6}\;\frac{{\theta}_2{\theta}_3{\theta}_4}
{({\theta}_2+{\theta}_3+{\theta}_4)^2}
\left[
({\theta}_3^4+{\theta}_4^4)
{\theta}_2^3 \ln{\frac{{\theta}_3}{{\theta}_4}}
-
({\theta}_2^4+{\theta}_3^4)
{\theta}_4^3\ln{\frac{{\theta}_2}{{\theta}_3}}
+
({\theta}_2^4-{\theta}_4^4)
{\theta}_3^3\ln{\frac{{\theta}_2}{{\theta}_4}}\right]
\nonumber \\
&-&\frac{\pi}{9}\;\frac{{\theta}_2^5+{\theta}_3^5
+{\theta}_4({\theta}_2^4+{\theta}_3^4)-2{\theta}_2{\theta}_3
({\theta}_2^3+{\theta}_3^3)}
{{\theta}_2+{\theta}_3+{\theta}_4}\left(1-2 \xi \theta_3^2 E\right)
\label{c2}
\end{eqnarray}
Equation (\ref{ratio}) imply that the amplitude of the term
$O(1/N^5)$ in Eq. (\ref{spheatfin}), i.e. $c_5$,
is $\sqrt{2} u_5$ where $u_5$ is the amplitude of the $N^{-5}$
correction terms in the internal energy
expansion Eq. (\ref{internfin}).

In two-eigenvalues approximation the specific heat can be written as
\begin{equation}
C=\frac{8}{\pi} \ln{N}+c_0^{(app)}+\frac{c_1^{(app)}}{N}+
\frac{c_2^{(app)}}{N^2}+\frac{c_3^{(app)}}{N^3}+...
\label{Capp}
\end{equation}
with
\begin{eqnarray}
c_0^{(app)}&=&\frac{8}{\pi}\left(\ln{\frac{2^{5/2}}{\pi}}+C_E
-\frac{\pi}{4}\right)+\xi \; {\rm sech}^2{\left(\pi \xi/8\right)}
+\frac{8 \ln{2}}{\pi}\left[-1+\tanh{\left(\pi \xi/8\right)}\right]
\label{c0app}\\
c_1^{(app)}&=&\sqrt{2}\left[-1+\tanh{\left(\pi \xi/8\right)}\right]
\label{c1app}\\
c_2^{(app)}&=& -\frac{\pi}{9}
+\frac{\pi}{6}\left[-1+\tanh{\left(\pi \xi/8\right)}\right]
+\frac{\pi^2 \xi \ln{2}}{24}{\rm sech}^2{\left(\pi \xi/8\right)}
\nonumber\\
&-&\frac{\pi^3 \xi^2}{96}{\rm sech}^2{\left(\pi \xi/8\right)}
\tanh{\left(\pi \xi/8\right)}
\label{c2app}\\
c_3^{(app)}&=& \frac{\pi^3 \xi}{96 \sqrt{2}}
{\rm sech}^2{\left(\pi \xi/8\right)}
\label{c3app}
\end{eqnarray}
We plot the aspect-ratio ($\xi$) dependence of the finite-size specific heat
correction terms $c_0$ and $c_2$ in Fig. 3. The exact and approximate values
approach exponentially as $\xi$ increases. Note, that the ratios of
correction terms $u_{1}/c_{1}$ and $u_{3}/c_{3}$  are constant and given
by Eq. (\ref{ratio}).
In Fig. 4 we plot the aspect-ratio dependence of the error introduced by
two-eigenvalue approximation for the correction terms in the free energy,
internal energy and specific heat asymptotic expansions.
The deviation of the two-eigenvalues approximation from exact result is
about one percentage at $\xi=1$ for the leading correction terms
$f_1, u_1, c_0$, at $\xi = 1.65$ for the second correction terms
$f_3, u_3, c_2$, at $\xi = 1.85$ for the third correction terms $f_5, u_5$
and diminishes very rapidly as $\xi$ increases.

It is of interest to compare this finding with other results.
Equations (\ref{freeenfin}), (\ref{internfin}), and (\ref{spheatfin}) are
consistent with Ferdinand and Fisher's similar expansions \cite{ff69} up to
orders $1/N^2$, $1/N$, and $1/N$, respectively. Others terms in our
equations, except the term of $O(1/N^3)$ for $U$ \cite{hcik99}, are new.
For $\xi=1$, we have $u_3=0.206683145\dots$ and $u_5=0.730182312347\dots$
which are quite consistent with numerical data
$u_3=0.206683133$ and $u_5=0.73018231235$ obtained by
 Salas and Sokal \cite{sokal}.

\vskip 8 mm

\section{Discussion}

The results of this paper inspire several problems for further
studies: (i) can one obtain an exact asymptotic expansion for the
thermodynamic functions up to arbitrary order, as it can be done for the
Ising model on $N \times \infty$ square, honeycomb, and plane triangular
lattices \cite{ih00}. (ii) It is of interest to know whether the
 amplitude ratio of Eq. (\ref{ratio})
can be extended to honeycomb and plane triangular lattices, i. e. whether
the ratio is universal.
 (iii) If, so, how do such amplitudes behave in
other models, for example in the three-state Potts model?

{\it Note added:} After the completion of this paper, we learned that similar
results have been independently obtained by Salas \cite{salas1}.

\vskip 8 mm
\section{Acknowledgments}

We thank J. Salas for helpful communications.
This work was supported in part by the National Science Council of the
Republic of China (Taiwan) under Contract No. NSC 89-2112-M-001-084.

\vskip 8 mm

\begin{appendix}
{\vskip 4 mm}
Let us now evaluate the coefficients $p_{i1}$ and $p_{i2}$ for
$i = 1, 2, 3, 4$. After little algebra, following the general lines
of the Ferdinand and Fisher paper \cite{ff69}, we can obtain the
following expression for the coefficients $p_{i1}$ and $p_{i2}$

\begin{eqnarray}
p_{11}&=&-
\frac{1}{3}\pi^3 \xi
\left[\frac{1}{4}\sum_{j=1}^{\infty}(-1)^{j}\frac{\cosh{\pi\xi j}}
{\sinh^2{\pi\xi j}}+\frac{3}{2}\sum_{j=1}^{\infty}(-1)^{j}
\frac{\cosh{\pi\xi j}}{\sinh^4{\pi\xi j}}\right]
\nonumber \\
p_{21}&=&-
\frac{1}{3}\pi^3 \xi
\left[\frac{1}{4}\sum_{j=1}^{\infty}\frac{\cosh{\pi\xi j}}{\sinh^2{\pi\xi j}}
+\frac{3}{2}\sum_{j=1}^{\infty}\frac{\cosh{\pi\xi j}}{\sinh^4{\pi\xi j}}
\right]
\label{pi1} \\
p_{31}&=&-
\frac{1}{3}\pi^3 \xi
\left[\sum_{j=1}^{\infty}\frac{(-1)^{j}}{\sinh^2{\pi\xi j}}
+\frac{3}{2}\sum_{j=1}^{\infty}\frac{(-1)^{j}}{\sinh^4{\pi\xi j}}\right]
\nonumber \\
p_{41}&=&-
\frac{1}{3}\pi^3 \xi \left[\sum_{j=1}^{\infty}\frac{1}{\sinh^2{\pi\xi j}}
+\frac{3}{2}\sum_{j=1}^{\infty}\frac{1}{\sinh^4{\pi\xi j}}\right]
\nonumber\\
p_{12}&=&\frac{1}{2}p_{11}^2+\frac{4 \pi^6 \xi^3}{81}\left[
\frac{3}{\pi \xi}\Psi_1(\pi\xi)+\frac{d}{d(\pi\xi)}\Psi_1(\pi\xi)\right]
\nonumber \\
p_{22}&=&\frac{1}{2}p_{21}^2+\frac{4 \pi^6 \xi^3}{81}\left[
\frac{3}{\pi \xi}\Psi_2(\pi\xi)+\frac{d}{d(\pi\xi)}\Psi_2(\pi\xi)\right]
\nonumber \\
p_{32}&=&\frac{1}{2}p_{31}^2+\frac{4 \pi^6 \xi^3}{81}\left[
\frac{3}{\pi \xi}\Psi_3(\pi\xi)+\frac{d}{d(\pi\xi)}\Psi_3(\pi\xi)\right]
\label{pi2} \\
p_{42}&=&\frac{1}{2}p_{41}^2+\frac{4 \pi^6 \xi^3}{81}\left[
\frac{3}{\pi \xi}\Psi(\pi\xi)+\frac{d}{d(\pi\xi)}\Psi(\pi\xi)\right]
\nonumber
\end{eqnarray}
with
\begin{eqnarray}
\Psi_1(x)&=&\frac{1}{128}\sum_{j=1}^{\infty}(-1)^j\cosh{x j}\left(
\frac{1}{\sinh^2{x j}}+\frac{60}{\sinh^2{x j}}
+\frac{120}{\sinh^6{x j}}\right)
\nonumber \\
\Psi_2(x)&=&\frac{1}{128}\sum_{j=1}^{\infty}\cosh{x j}\left(
\frac{1}{\sinh^2{x j}}+\frac{60}{\sinh^4{x j}}
+\frac{120}{\sinh^6{x j}}\right)
\nonumber \\
\Psi_3(x)&=&\frac{1}{16}\sum_{j=1}^{\infty}(-1)^j\left(
\frac{2}{\sinh^2{x j}}+\frac{15}{\sinh^4{x j}}
+\frac{15}{\sinh^6{x j}}\right)
\label{Psi} \\
\Psi(x)&=&\frac{1}{16}\sum_{j=1}^{\infty}\left(
\frac{2}{\sinh^2{x j}}+\frac{15}{\sinh^4{x j}}
+\frac{15}{\sinh^6{x j}}\right)
\nonumber
\end{eqnarray}
Let us now introduce the following notation:
\begin{equation}
S_n(x)=\sum_{j=1}^{\infty}\frac{1}{\sinh^n{x j}} \qquad \mbox{for}
\qquad n = 2, 4, 6.
\label{Snx}
\end{equation}
Then the coefficients $p_{i1}$ and $p_{i2}$  can be
rewritten in the more symmetrical way
\begin{eqnarray}
p_{11}&=&
\frac{1}{3}\pi^3\xi\left[\frac{1}{8}R(\xi/2)
-\frac{5}{4} R(\xi)+2 R(2\xi)\right]
\nonumber \\
p_{21}&=&
\frac{1}{3}\pi^3\xi\left[R(\xi)-\frac{1}{8}R(\xi/2)
\right]
\label{Ri} \\
p_{31}&=&
\frac{1}{3}\pi^3\xi [R(\xi)-2 R(2\xi)]
\nonumber \\
p_{41}&=&
-\frac{1}{3}\pi^3\xi R(\xi)
\nonumber
\end{eqnarray}
and
\begin{eqnarray}
\Psi_1(\xi)&=&-2 \Psi(2 \xi)+\frac{17}{16}\Psi(\xi)-\frac{1}{32}\Psi(\xi/2)
\nonumber\\
\Psi_2(\xi)&=&-\Psi(\xi)+\frac{1}{32}\Psi(\xi/2)
\label{Psii}\\
\Psi_3(\xi)&=&2 \Psi(2 \xi)-\Psi(\xi)
\nonumber
\end{eqnarray}
with
\begin{eqnarray}
R(x)&=&S_2(x) + \frac{3}{2} S_4(x)
\label{Rx}\\
\Psi(x)&=&\frac{1}{16}\left(2 S_2(x)+15 S_4(x)+15 S_6(x)\right)
\label{Psix}
\end{eqnarray}
Thus we have shown that the coefficients $p_{i1}$ and $p_{i2}$  can be
expressed in terms of the only object, namely $S_n(x)$ for $n=2, 4, 6$.
The $S_2(x)$ is given by (see \cite{prud} p. 721)
\begin{equation}
S_2(x)=
\frac{1}{6}+
\frac{2(2-k^2)}{3 {\pi}^2}K^2(k)-\frac{2}{{\pi}^2}K(k)E(k)
\label{S2x}
\end{equation}
where $x=\pi K'(k)/K(k)$ with $K(k)$ and $E(k)$ the elliptic integrals of the
first and second kind, respectively.
The $S_4(x)$ and $S_6(x)$ are calculated to be
\begin{eqnarray}
S_4(x)&=&
-\frac{11}{90}-\frac{4(2-k^2)}{9 {\pi}^2}K^2(k)+\frac{4}{3 \pi^2}K(k)E(k)
\label{S4x} \\
&+&
\frac{8(1-k^2+k^4)}{45 {\pi}^4}K^4(k)
\nonumber \\
S_6(x)&=&\frac{191}{1890}+\frac{32(2-k^2)}{45 \pi^2}K^2(k)
-\frac{16}{15 \pi^2}K(k)E(k)
\label{S6x}\\
&-&\frac{8(1-k^2+k^4)}{45 \pi^4}K^4(k)-
\frac{32(2-3 k^2-3 k^4+2 k^6)}{945 \pi^6}K^6(k)
\nonumber
\end{eqnarray}
Thus we are now in position to evaluate $R(x)$ and $\Psi(x)$ given by
Eqs. (\ref{Rx}) and  (\ref{Psix}) respectively. The result is
\begin{eqnarray}
R(x)&=&-\frac{1}{60}+\frac{4(1-k^2+k^4)}{15\pi^4}K^4(k)
\label{R1} \\
\Psi(x)
&=&\frac{1}{1008}-\frac{2(2-3 k^2-3 k^4+2 k^6)}{63 \pi^6}K^6(k)
\label{Psi1}
\end{eqnarray}
The expressions for $R(2 x), R(x/2)$ and $\Psi(2 x), \Psi(x/2)$ can be
written as function of the modulus $k$ by using properties of the elliptic
functions.

Thus for the coefficients $p_{i1}$ (for $i = 1, 2, 3, 4$) we
have finally obtained
\begin{eqnarray}
p_{11} &=& -\frac{7\pi^3 \xi}{1440}
+ \frac{(7+8 k^2-8 k^4)\xi}{90 \pi} K^4(k),
\nonumber \\
p_{21}&=&-\frac{7\pi^3 \xi}{1440}
+ \frac{(7-22 k^2+7 k^4)\xi}{90 \pi} K^4(k),
\label{pi1fin} \\
p_{31}&=& \frac{\pi^3 \xi}{180}
+ \frac{(-8+8 k^2+7 k^4)\xi}{90 \pi} K^4(k),
\nonumber \\
p_{41}&=& \frac{\pi^3 \xi}{180}
- \frac{4 (1-k^2+ k^4)\xi}{45 \pi} K^4(k).
\nonumber
\end{eqnarray}
After little algebra the expressions for $p_{i2}$ (for $i = 1, 2, 3, 4$)
can be written as
\begin{eqnarray}
p_{12} &=&
\frac{\xi^2 K^8}{189 \pi^2}\left[31-88 k^2+88 k^4+
(31-78 k^2+48 k^4-32 k^6)\left(\frac{E'}{K'} - \frac{E}{K}\right)
\right]
\nonumber \\
&-&\frac{31 \pi^5 \xi}{24192}+\frac{p_{11}^2}{2}
\nonumber\\
p_{22}&=&\frac{\xi^2 K^8}{189 \pi^2}\left[31-67 k^2+25 k^4-21 k^6+
(31-15 k^2-15 k^4+31 k^6)\left(\frac{E'}{K'} - \frac{E}{K}\right)
\right]
\nonumber\\
&-&\frac{31\pi^5 \xi}{24192}+\frac{p_{21}^2}{2}
\label{pi2fin} \\
p_{32}&=&-\frac{\xi^2 K^8}{189 \pi^2}\left[32-80 k^2+38 k^4-21 k^6+
(32-48 k^2+78 k^4-31 k^6)\left(\frac{E'}{K'} - \frac{E}{K}\right)
\right]
\nonumber\\
&+&\frac{\pi^5 \xi}{756}+\frac{p_{31}^2}{2}
\nonumber \\
p_{42}&=& -\frac{16 \xi^2 K^8}{189 \pi^2}\left[2-5 k^2+5 k^4+
(2-3 k^2-3 k^4+2 k^6)\left(\frac{E'}{K'} - \frac{E}{K}\right)
\right]
\nonumber\\
&+&\frac{\pi^5 \xi}{756}+\frac{p_{41}^2}{2}
\nonumber
\end{eqnarray}
where for simplicity we denote $K \equiv K(k)$, $K' \equiv K'(k)$,
$E \equiv E(k)$ and $E' \equiv E'(k)$.
\end{appendix}

\newpage

\begin{center}
{\bf FIGURE CAPTIONS}
\end{center}
\vskip 0.6cm

{\bf FIG. 1.} Finite-size free energy correction terms (a) $f_1$, (b) $f_3$
and (c) $f_5$ as functions of the aspect ratio $\xi$, which are
defined by Eqs. (37), (38), (39), and (43). Solid curves: exact values; dashed
 curves: two-eigenvalue approximations of Eqs. (49)-(51).
The exact and approximate values approach exponentially as $\xi$ increases.

\vskip 0.4cm

{\bf FIG. 2.}  Finite-size internal energy correction
terms (a) $u_1$, (b) $u_3$ and (c) $u_5$ as functions of
the aspect ratio $\xi$, which are defined by Eqs. (40)-(43).
Solid curves: exact values; dashed curves:
two-eigenvalue approximations of Eqs. (52)-(54).
The exact and approximate values approach
exponentially as $\xi$ increases.

\vskip 0.4cm

{\bf FIG. 3.} Finite-size specific heat correction terms (a) $c_0$
and (b) $c_2$ as functions of the aspect
ratio $\xi$, which are defined by Eqs. (64) and (65).
Solid curves: exact values; dashed curves: two-eigenvalue
approximations of Eqs. (67) and (69). The exact and approximate
values approach exponentially as $\xi$ increases.

\vskip 0.4cm

{\bf FIG. 4.} The error introduced by two-eigenvalue
approximation for the correction terms in the free energy,
internal energy and specific heat asymptotic expansions.
(a) The error for the
leading correction terms $f_1$, $u_1$ and $c_0$ as functions of $\xi$.
(b) The error for the second correction terms
$f_3$, $u_3$ and $c_2$  as functions of $\xi$.
(c) The error for the third correction
terms $f_5$ and $u_5$ as functions of $\xi$.
The vertical axes represents the error
defined by $\mbox{Error}=\left(a^{exact}-a^{approx}\right)/a^{exact}$, where
$a$ stand for the correction terms in the free energy, internal energy and
specific heat asymptotic expansions.
Solid curves: free energy; dashed curves: internal energy;
dot-dashed curves: specific heat.  The deviation of the
two-eigenvalues approximation from exact result is about one percentage at
$\xi=1$ for the leading correction terms, at
$\xi = 1.65$ for the second correction terms, at $\xi = 1.85$
for the third correction terms and diminishes
very rapidly as $\xi$ increases.

\end{document}